\documentclass[numbers,preprint,endfloats]{jasatex}

\usepackage[dvips]{color}
\usepackage{setspace}
\usepackage{mathpazo}

\usepackage{graphicx}
\usepackage{amsmath}
\usepackage{amssymb}
\usepackage{amsfonts}
\usepackage{subfigure}
\usepackage{float}
\usepackage{url}
\usepackage{booktabs}
\usepackage{gensymb}

   \newcommand{\beq}{\begin{equation}}
   \newcommand{\eeq}{\end{equation}}

\begin{document}

\title{Interacting bubble clouds and their sonochemical production}




\author{
Laura Stricker\\Physics of Fluids Group, Impact Institute \&
Burgers Center for Fluid Dynamics, \\University of Twente, \\Drienerlolaan 5, \\7500 AE Enschede, The Netherlands \\[\baselineskip]
Benjamin Dollet \\
Institut de Physique de Rennes, \\ UMR 6251 CNRS/Universit\'e de
Rennes 1, \\B\^atiment 11A, Campus Beaulieu, Avenue du
G\'{e}n\'{e}ral Leclerc 263\\ 35042 Rennes Cedex,
France \\[\baselineskip]
David Fern\'{a}ndez Rivas\\
Mesoscale Chemical Systems Group, MESA+ Research Institute, \\
University of Twente, \\ Drienerlolaan 5, \\ 7500 AE Enschede, The
Netherlands \\[\baselineskip]
Detlef Lohse \\
Physics of Fluids Group,  Impact Institute \& Burgers Center for
Fluid Dynamics, University of Twente,  \\Drienerlolaan 5, \\
7500 AE Enschede, The Netherlands}

\date{\today}

\begin{abstract}
Acoustically driven air pockets trapped in artificial crevices on
a surface can emit bubbles which organize in (interacting) bubble
clusters. With increasing driving power Fern\'{a}ndez Rivas
\emph{et al.} [Angew. Chem. Int. Ed., 2010] observed three
different behaviors: clusters close to the very pits out of which
they had been created, clusters pointing toward each other, and
merging clusters. The latter behavior is highly undesired for
technological purposes as it is associated with a reduction of the
radical production and an enhancement of the erosion of the
reactor walls. The dependence on the control parameters such as
the distance of the pits and the conditions for cluster-merging
are examined. The underlying mechanism, governed by the secondary
Bjerknes forces, turns out to be strongly influenced by the
nonlinearity of the bubble oscillations and not directly by the
number of nucleated bubbles. The Bjerknes forces are found to
dampen the bubble oscillations, thus reducing the radical
production. Therefore, the increased number of bubbles at high
power could be the key to understand the experimental observation
that, after a certain power threshold, any further increase of the
driving does not improve the sonochemical efficiency.
\end{abstract}

\pacs{4335Vz, 4335Ei}

\maketitle

\section{Introduction}
Sonochemistry is the use of ultrasound to achieve a chemical
conversion. Imploding microbubbles can produce localized extreme
temperature and pressure conditions. As a result, high energy
chemical conversions can be triggered, eventually resulting in the
production of highly reactive radical species
\cite{sus90b,sus90c,sus99,cru99,mas02,sus08}. The applications of
these reaction products are manifold, including synthesis of fine
chemicals, food ingredients and pharmaceuticals, degradation of
water contaminants \cite{che91,kot91,kot92,gog01}, textile
processing \cite{yec99} and cell disruption \cite{shi98}.
Fern\'{a}ndez Rivas \emph{et al.} \cite{fer10,fer13} have recently
proposed an efficient way to do sonochemistry, controlling
cavitation by using micropits grooved on silicon substrates,
following the original idea of Bremond \emph{et al.} \cite{bre06}.
At sufficient acoustic pressures, a bubble cluster is generated in
the liquid above each pit. If two pits or more are present, such
clusters tend to attract and merge over a certain pressure
amplitude. It has been shown that merging clusters are associated
to a reduction in the radical production respect to the case of
not interacting clusters \cite{fer13} and erosion of the reactor
walls \cite{fer12b}. Therefore, in efficient sonochemical reactors
design, one should maximize the number of pits (i.e. of bubbles)
but should avoid cluster-merging conditions. The goal of the
present work is to understand the transition between the three
possible behaviors observed in experiments (see
Fig.~\ref{Fig:clusters behavior, experiment}): individual clusters
next to the pit out of which they were generated (behavior~1),
individual clusters pointing towards the central array of the pits
(behavior~2), and clusters migrating towards the center
(behavior~3). The key factor to study these phenomena are the
acoustic interactions between different bubbles, namely the
secondary Bjerknes forces \cite{bje06,lei94}.

 Though these forces have been largely investigated both for
bubble pairs \cite{bla49,cru75,ogu90,pel93,pel93b,doi99} and for
bubble clouds \cite{doi04,yas08}, making an a-priori prediction
even on their sign is a non-trivial matter. The linear theory
predicts that two acoustically driven bubbles oscillate in phase
and attract each other when the driving frequency is greater or
lower than both their resonance frequencies, while they oscillate
in counter-phase and repel with a driving frequency in-between
their resonance frequencies \cite{pro84b}. Thus, Bjerknes forces
are expected to be attractive for bubbles of equal size
\cite{pel93,pel93b}. As a first, qualitative statement, we can
therefore expect the cluster-cluster interaction to be attractive
(since the pits are identical, the bubbles of each cluster should
have similar sizes). Hence, the fact that clusters merge only
above a certain threshold suggests that the pit-cluster
interaction is also attractive, and that the cluster-cluster
interaction must overcome the pit-cluster one to achieve cluster
merging. However, it has been proved by a number of authors that
the sign of the Bjerknes forces can be reversed due to several
mechanisms neglected by the classical linear theory, such as the
effect of secondary harmonics \cite{ogu90,doi99}, the
resonance-like behavior of small bubbles (below their resonance
size) near the dynamic Blake threshold \cite{met97} and viscous
effects during translational motion \cite{doi02}. Several studies
of two bubbles interacting in a strong acoustic field have also
shown that bubbles oscillating nonlinearly can form a bound pair
with a steady spacing rather than collide and coalesce, as linear
Bjerknes theory would predict \cite{doi01,har01,pel04,yos11}.
Therefore caution is required when one wishes to understand the
behavior of interacting clusters of bubbles.

In the present work we will investigate the Bjerknes forces acting
upon the clusters and their dependence on different parameters,
such as the size and the number of the bubbles, the size of the
clusters, the distance between the pits and the driving pressure.
We will show the influence of these forces upon the bubble
dynamics and the radical production and will address the
conditions required for the different transitions, thus providing
practical indications for efficient sonochemical sonoreactors
design, where one wishes to have the highest possible number of
non-interacting micropits.

\section{Model} \label{sec:2}

In the experiment of Ref.~\cite{fer13}, the bubble population is
quite polydisperse. In top view, clusters appear as diffuse
circles of radius $R_c$ (Fig.~\ref{Fig:clusters behavior,
experiment}); a side view reveals that they are in contact with
the substrate and have roughly an elliptic shape (Fig.~7 of
Ref.~\cite{fer13}) with a long axis parallel to the substrate, and
not much larger than the short axis. To simplify the problem, we
will assume that each cluster is a sphere of radius $R_c$ in
tangential contact with the substrate and we will neglect
polydispersity. We will consider $R_c$ as time-invariant for
simplicity reasons, although in reality this holds only on
average, for a give applied power, but not in a rigorous sense. We
will also assume that all the bubbles have the same radius $R(t)$
and that the two cluster have the same number of bubbles $N$,
constant in time.

Each individual bubble in a cluster experiences acoustic
interactions from the pits, from the neighboring bubbles of the
same cluster, from the bubbles of the other cluster, and from all
image bubbles, given the presence of the hard silicon substrate
(see Fig.~\ref{Fig:Sketch forces}). We describe the behavior of a
single bubble belonging to one of the clusters by extending the
model previously developed in Refs. \cite{toe00,toe02,toe03} and
validated in \cite{str11}, to incorporate the effect of the
secondary Bjerknes forces upon the pulsation of the bubble. This
model is then coupled with a static force balance in order to
study the switch between the different conditions observed
experimentally (see Fig.~\ref{Fig:clusters behavior, experiment}).

The model that we adopt for the single bubble is an ODE model
based on the assumptions that the gas inside the bubble is a
perfect gas and that the bubble has a uniform temperature and
pressure. The temperature evolution is derived from the energy
equation. Heat and mass transfer are treated with a boundary layer
approximation \cite{toe00,toe02,toe03}. Evaporation/condensation
phenomena are kept into account, as well as the variation of the
transport parameters due to compositional changes of the mixture.
A list of 45 chemical reactions is included, with their
temperature dependent chemical kinetics, governed by Arrhenius
law. We refer the reader to \cite{toe03} for a detailed
description of these parts of the model, and we concentrate in the
following upon the treatment of the Bjerknes forces.

The radial dynamics of a bubble belonging to cluster~1 is
described by means of a modified Rayleigh-Plesset equation,
keeping into account the effect of the secondary Bjerknes forces
on the radial pulsation \cite{yas08,zer11}: \beq
\begin{split} \left(1 - \frac{\dot{R}}{c}\right) &R\ddot{R}+
\frac{3}{2} \left(1-\frac{\dot{R}}{3c}\right) \dot{R}^{2} \\
 = \frac{1}{\rho} &\left(1+\frac{\dot{R}}{c}\right)
 \left(p - p_{\infty} - P(t) \right) +
 \frac{R \dot{p}}{\rho c} - \frac{4 \nu \dot{R}}{R} -
 \frac{2\sigma}{\rho R} - T_{Bj} \, .
 \end{split} \label{Eq:modif R-Plesset eqn} \eeq
Here the dots are used for time derivatives, $R$ is the radius of
the bubble, $c$ is the speed of sound, $\rho$ is the density of
the liquid, $\nu$ its kinematic viscosity, $\sigma$ is the surface
tension, $p_\infty$ the static pressure and $P(t)=P_a \cos\omega t
$ is the acoustic driving pressure, with $P_a$ the driving
amplitude, $f = \omega / 2\pi$ the frequency and $\tau = 1/f$ the
period of the driving. $T_{Bj}$ is a coupling term expressing the
effect of the interaction with the other bubbles, both real and
imaginary, and the pits \beq T_{Bj} = T_{c1} + T_{c2} + T_{m1} +
T_{m2} + T_{p1} + T_{p2} \, . \label{coupling terms}\eeq $T_{c1}$
and $T_{c2}$ are the coupling terms with the bubbles of the same
cluster and the other cluster respectively, $T_{m1,2}$ are the
coupling terms with the two mirror clusters, $T_{p1,2}$ are the
coupling terms with the two pits.

The coupling term $T_{2\rightarrow 1}$ between two isolated
bubbles, describing the influence of bubble~2 on the radial
oscillations of bubble~1, can be written as \cite{met97} \beq
T_{2\rightarrow 1} = \frac{1}{d} (R_2^2 \ddot{R}_2 + 2R_2
\dot{R}_2^2) \, . \label{Eq:Tbj_bub2,bub1}\eeq Therefore, the
coupling term between one bubble $i$ and the other bubbles of the
cloud to which it belongs becomes \beq T_{c1}= \sum_{j\neq i}
\frac{R_j^2 \ddot{R}_j + 2R_j \dot{R}_j^2}{r_{ij}} \, ,
\label{Eq:Tbj_cloud1,bub1 generic}\eeq in which $r_{ij}$ is the
distance between bubbles $j$ and $i$. Following the approximation
of Yasui \emph{et al.} \cite{yas08}, i.e. neglecting
polydispersity and assuming that the cluster has constant density,
we get: \beq T_{c1} \simeq 2\pi nR_c^2 (R^2 \ddot{R} +
2R\dot{R}^2) \, ,\label{Eq:Tbj_cloud1,bub1}\eeq where $R_c$ is the
cluster radius and $n$ is the number density of bubbles, $n \simeq
3N/4\pi R_c^3$.

The coupling term between the considered bubble of cluster~1 and
all the bubbles of cluster~2 is expressed as in Ref. \cite{yas08}
\beq T_{c2} \simeq \frac{N}{d_{c}} (R^2 \ddot{R} + 2R\dot{R}^2) \,
, \label{Eq:Tbj_cloud2,bub1} \eeq where $d_c$ is the distance
between the two clouds. This is valid as long as $d_c \gg R_c$.
The coupling terms $T_{m1,2}$ with the two mirror clusters are
expressed in a similar way.

In order to model the interaction of the bubble with the pit, we
consider the pit as an effective bubble with the same resonant
frequency $\omega_p$ and damping coefficient $\beta$ as the pit.
Although the pit is not a spherical bubble, and may oscillate in a
nonlinear fashion, we will treat it as a harmonic oscillator. This
assumption is acceptable as long as we showed in
Ref.~\cite{str12_thesis} that, in the considered parametric range,
large amplitude oscillations of a gas pocket entrapped inside a
cylindrical pit present an overall behavior similar to small
amplitude oscillations, with a slightly lower damping but the same
resonance frequency. Hence, we consider the pit like a spherical
bubble of radius at rest $R_p^0$, such that $\omega_p^2 =
3p_\infty/\rho (R_p^0)^2$, experiencing linear oscillations
\cite{pro84}: $R_p = R_p^0 (1 + x_p)$ with:
\begin{equation}\label{Eq:linear_xp}
\ddot{x}_p + 2 \beta\dot{x}_p + \omega_p^2 x_p =
-\frac{p_a(t)}{\rho (R_p^0)^2} .
\end{equation}

The values of $\omega_p$ and $\beta$ are taken from the recent
results of Gelderblom \emph{et al.} \cite{gel12}. These authors
have computed the acoustic response of a gas pocket entrapped in a
pit, providing its eigenfrequency and damping coefficient, in two
limits: potential flow, and unsteady Stokes equation. For a
cylindrical pit of radius $a$ and height $h$, the results mainly
depend on the parameter $P = \kappa a^2 p_0^\infty/h\sigma$. In
the experiments of \cite{fer10,fer13}, $a = 15~\mu$m, $h =
10~\mu$m, $p_0^\infty = 10^5$~Pa and $\sigma = 0.07$~N/m. The
temperature was controlled providing an isothermal behavior within
a precision of 1~K \cite{fer13} and therefore $\kappa = 1$. Hence
we compute $P = 32$, from which $\hat{\omega}_p = 5.82$ and
$\hat{\beta} = 0.26$ in the Stokes regime, and $\hat{\omega}_p =
6.04$ and $\hat{\beta} = 0.20$ in the potential regime. The
dimensionless frequency is defined as $\hat{\omega}_p = \omega_p
\sqrt{\rho a^3/\sigma}$, with a rescaling angular frequency
$\sqrt{\sigma/\rho a^3} = 1.46\times 10^5$~rad/s. Hence, taking
$\hat{\omega}_p = 6$, the resonance frequency of a pit equals
$f_p$ = 143~kHz, with $f_p=\omega_p /2\pi$. In the experiments of
\cite{fer10,fer13}, it is not clear which of the two regimes,
potential or Stokes, applies best, but it is seen that the
numerical values of eigenfrequency and damping differ by only
10\%. Moreover, in the numerical work of Ref.~\cite{str12_thesis}
it was shown that in the intermediate regime where both inertia
and viscosity are present, the overall behavior of the pit is
closer to the Stokes regime, when the static pressure equals
$p_\infty$ = 1~atm. So, we use hereby the results related to the
Stokes regime for the damping and we take $\beta=3.8 \times
10^5$~s$^{-1}$.

By substituting $R_p(t)$ inside (\ref{Eq:Tbj_bub2,bub1}), the
coupling term between the considered bubble of cluster~1 and pit~1
can be expressed as \beq T_{p1} =
\frac{1}{d_{p1}}(R_p^0)^3(1+x_p)(2\dot{x}_p^2+x_p\ddot{x}_p+\ddot{x}_p)
\, , \label{Eq:Tbj_pit1,bub1} \eeq where $d_{p1}$ is the distance
between the bubble and the pit. The coupling term $T_{p2}$ between
the bubble of cluster~1 and pit~2 is expressed correspondingly.

We now turn to the forces experienced by the clusters (see
Fig.~\ref{Fig:Sketch forces}). As we want to study the transition
between behavior~1 and behavior~2, we will focus on the forces
acting on the horizontal plane. Hence, we will neglect both the
buoyancy and the primary Bjerknes force, which is directed in the
vertical direction, because the driving pressure is a standing
wave with a anti-node on the substrate and a node at the free
water-air surface \cite{zij11}.

In a static equilibrium condition, all the forces acting upon a
bubble of cluster~1 with an horizontal component are secondary
Bjerknes forces. Such forces are exerted by the other cluster
$\vec{F}_c$, by the mirror clusters $\vec{F}_{m1,2}$ and by the
pits $\vec{F}_{p1,2}$. In order to derive them, we first consider
two oscillating bubbles, of volume $\mathcal{V}_1$ and
$\mathcal{V}_2$, separated by a distance $d$ much greater than
their radii; then bubble~1 experiences a force equal to
\cite{met97}:
\begin{equation}\label{Eq:secondary_Bjerknes_force}
\vec{F}_{2\rightarrow 1} = \frac{\rho\ddot{\mathcal{V}}_2
\mathcal{V}_1}{4\pi d^2} \hat{e}_{2\rightarrow 1} ,
\end{equation}
with $\hat{e}_{2\rightarrow 1}$ the unit vector pointing from
bubble~2 to bubble~1. For bubbles of the same radius $R$, this
reduces to $\vec{F}_{2\rightarrow 1} = 4\pi\rho R^3 (R^2 \ddot{R}
+ 2R\dot{R}^2) \hat{e}_{2\rightarrow 1}/3d^2$. Let us first notice
that the forces between bubbles pertaining to the same cluster are
responsible of the cluster cohesion, but are irrelevant to the
interaction between different clusters; hence, we neglect them to
assess the stability of behavior~1. Assuming that $R_c \ll d_c$,
each bubble of cluster~1 experiences from cluster~2 a force equal
to \cite{yas08,zer11}:
\begin{equation}\label{Eq:cluster_cluster_force}
\vec{F}_c = \frac{4\pi\rho NR^3 (R^2 \ddot{R} +
2R\dot{R}^2)}{3d_c^2} \hat{e}_{2\rightarrow 1} .
\end{equation} As $d_c = d - 2\delta$
(Fig.~\ref{Fig:Sketch forces}), the horizontal component of
 $\vec{F}_c$ is given by \beq \label{Eq:Fc(delta)} F_{c,x} = \frac{A_c}{(d - 2\delta)^2}  \, , \eeq
 with $A_c = 4\pi\rho N R^3 (R^2 \ddot{R} + 2R\dot{R}^2)/3$.
 The forces acting on each bubble of cluster~1 from the mirror clusters, namely
$\vec{F}_{m1}$ and $\vec{F}_{m2}$, and their horizontal components
can be expressed in a similar way.

The secondary Bjerknes force acting over each bubble of cluster~1
from pit~1 is found by substituting the volume of the equivalent
bubble corresponding to the pit $\mathcal{V}_p = 4\pi R_p^3/3$
into Eq.~(\ref{Eq:secondary_Bjerknes_force}),
\begin{equation}\label{Eq:pit_cluster_force}
\vec{F}_{p1} = \frac{4}{3} \frac{\pi\rho (R_p^0)^3 }{d_p^2} R^3
(1+x_p)\left[ 2\dot{x}_p^2 + \ddot{x}_p(1+x_p) \right] \hat{e}_{p}
\, .
\end{equation}
Here $\hat{e}_{p}$ is the unit vector pointing from the pit to the
bubble and $d_{p}^2 = h_p ^2 + \delta^2$, with $h_p = R_c + h/2$.
Even if $d_p$ depends on the location of the bubble within its
cluster, we will take $d_p$ as the distance between the pit and
the center of the cluster; in practice, off-centered bubbles
within the cluster will experience pit interaction of a different
magnitude, but this will be compensated by the interaction with
the other bubbles responsible of the cohesion of the cluster.
Given $\hat{e}_p \cdot \hat{e}_x = \delta/d_{p1}$, the horizontal
component of $F_{p1}$ can be calculated from
(\ref{Eq:pit_cluster_force}) as \beq \label{Eq:Fp(delta)} F_{p1,x}
= A_{p1} \frac{\delta}{[h_p^2 + \delta^2]^{3/2}} \, , \eeq with
$A_{p1} = 4\pi\rho(R_p^0)^3 R^3 (1+x_p)\left[ 2\dot{x_p}^2 +
\ddot{x_p}(1+x_p) \right]/3$. The force $F_{p2}$ acting on each
bubble of cluster~1 from pit~2 and its horizontal component are
calculated similarly.

\section{Results} \label{sec:3}

\subsection{Transition from individual clusters to merging clusters}

 Fern\'{a}ndez Rivas et al. \cite{fer13} provide an
experimental characterization of the number and size of the
bubbles present in the cluster for one, two and three pits, and
for three different powers. They show that the number of bubbles
$N$ increases at increasing power, and that both the average
bubble radius ($R\simeq$ 10~$\mu$m) and the most probable radius
($R\simeq$ 3~$\mu$m) have no significant dependence on the power
and on the number of pits. According to
(\ref{Eq:cluster_cluster_force}) and (\ref{Eq:pit_cluster_force}),
at first order, i.e. neglecting the effect of the other bubbles on
$R(t)$, the cluster-cluster force is proportional to $N$, whereas
the pit-cluster force does not depend on $N$. The threshold
pressure for merging may thus originate either from the fact that
$N$ increases at increasing power or from higher order effects of
bubble oscillations on the secondary Bjerknes forces, contained
inside the terms $A_p$ and $A_c$.
 In order to investigate the transitions between the
three different behaviors found in experiments (see
Fig.~\ref{Fig:clusters behavior, experiment}), we adopt a
quasi-static "adiabatic" approach: we take $\delta$ as a constant
over time, representing the displacement of the bubble and
therefore of the cloud from the initial equilibrium position. At
each instant, we calculate the horizontal components of
$\vec{F}_c$, $\vec{F}_{p1,2}$ and $\vec{F}_{m1,2}$. We stop the
calculation after one cycle, to match the experimental conditions
of Ref.~\cite{fer13}, where the bubbles did not survive after the
first collapse. We perform the time averages of the forces over
the whole cycle and we verify whether the following holds \beq
\langle F_x^* \rangle > \langle F_{p1,x}\rangle \, ,
\label{Eq:transition}\eeq where $\langle F_x^*\rangle = \langle
F_c\rangle + \langle F_{p2,x}\rangle + \langle F_{m2,x}\rangle$ is
the sum of the forces attracting the clusters towards each other
and $\langle \cdot \rangle$ denotes the time average over the
first acoustic cycle.

In order to study the transition between behavior~1 and
behavior~2, i.e. the inception of the motion, we consider a small
initial horizontal displacement $\delta = R_0$ of the bubble from
the pit axes and therefore from its rest conditions. The motion
starts once Eq.~(\ref{Eq:transition}) holds. However, this does
not imply that the bubble will eventually reach the center of the
pit array, as we will show below. An example of forces acting upon
a bubble with an initial displacement of $\delta = R_0$ is given
on Fig.~\ref{Fig:forces, delta=R0}. Sign inspection shows that the
cluster-cluster forces both with the real and the mirror cluster~2
are always attractive, while the pit-clusters forces become
repulsive at high driving pressure. Above pressures of 150~kPa the
noise increases, probably due to the nonlinearity of the problem.

In order to study the transition between behavior~2 and
behavior~3, i.e. the merging of the clusters, we consider both
$F_x^*$ and $F_{p1,x}$ as a function of $\delta$. The transition
occurs once Eq.~(\ref{Eq:transition}) holds at all $\delta$. As a
function of $\delta$, $\langle F_{p1,x}\rangle$ has a maximum,
generally (but not always) corresponding to the point
$\overline{\delta}$ where the pit traps the cluster. The
transition is graphically sketched on Fig.~\ref{Fig:transition}.
The black lines correspond to merging clusters (behavior~3); the
lit-gray lines correspond to the situation where $\langle F_x^*
\rangle$ is strong enough to induce the inception of the motion
but too weak to overcome the barrier constituted by the restoring
force of the pit. Therefore the cluster remains attached to its
pit, with a small displacement given by the intersection of the
two curves (behavior~2). The mid-gray lines represent the
transition between behavior~2 and behavior~3, and the
corresponding pressure amplitude will be denoted from now on as
$\bar{P_a}$.

In Fig.~\ref{Fig:d-Pa, transition} we plot the driving pressure
required for the two transitions, namely the one for cluster~1 to
start moving (dash-dotted line) and the one to overcome the
trapping force of the pit (solid line), as a function of the
distance between the pits $d$. According to the experimental
conditions of Ref.~\cite{fer13}, we consider two clusters with
$R_c$ = 100~$\mu$m, $N=100$, $R_0$ = 10~$\mu$m. As the distance
between the pits increases, a higher $P_a$ is required for
merging. For $d$ = 1000~$\mu$m, we calculated that the transition
occurs at $\bar{\delta}$ = 94~$\mu$m, with $\bar{P_a}$ = 270~kPa.
In the experiments, the maximum displacement of the clusters
before they detach from the pit and started coalescing was
$\bar{\delta}\sim R_c$ (see Fig.~\ref{Fig:clusters behavior,
experiment}). Moreover, three pressure values have been measured,
corresponding to the three different levels of of the applied
power (low, medium and high), 165~kPa, 225~kPa and 350~kPa
respectively. The transition occurred between 225~kPa and 350~kPa
\cite{fer13}. Despite the approximations of the model, such as the
equal size and monodispersity of the bubbles inside the cluster
and the time-invariance of both the number of the bubbles and the
size of the clusters, we remark that these values are extremely
close to the experimental ones.

Also as a function of the distance between the pits, the numerical
results reproduce the same trend found in experiments: when $d$
increases, so does $\bar{P_a}$, until a limiting value of $d$,
where no merging is possible anymore. In the experiments, this
limit was found at $d$ = 1500 $\mu$m, in the simulation at $d$ =
1350 $\mu$m. Once again, the agreement with the model is
remarkably good. The trapping distance $\bar{\delta}$ increases
with $d$. In Fig.~\ref{Fig:transition experiment}, we show the
voltage applied to the piezo when the clusters merge, at
increasing applied power (black) and when the clusters detach,
coming back to their own pits, at decreasing applied power (gray),
for the same setup and experimental conditions of
Ref.~\cite{fer13}. As the applied voltage is directly connected to
the driving pressure $P_a$, we can conclude that the driving
pressure required for merging is slightly higher than the one
required to detach the clusters. The theoretical investigation of
this hysteretic behavior is beyond the scope of the present paper
and should be addressed in future works.

For a given $R_0$ and $R_c$, $\bar{P_a}$ has a very slight
dependence on the number of bubbles and increasing $N$ reduces
$\bar{P_a}$ only until a certain value of $N$. With $R_0$ = 10
$\mu$m, $R_c$ = 100 and $d$ = 1000 $\mu$m this happens for $N <
50$ (see Fig.~\ref{Fig:transition, different N}). For clusters
with a higher number of bubbles, a further increase of $N$ does
not imply a further decrease of $\bar{P_a}$. This means that the
number of bubbles itself is not what determines the transition
between separated clusters and merging clusters. Thus, we can
conclude that, at medium and high power, the phenomenon is
governed by nonlinear oscillations effects contained inside $A_p$
and $A_c$ in Eqs.~(\ref{Eq:Fc(delta)}) and (\ref{Eq:Fp(delta)}).

However, the number of bubbles can still have an indirect
influence on the transition:  given a certain $R_0$ and $N$, both
$\bar{P_a}$ and $\bar{\delta}$ are higher when the cloud radius
$R_c$ is smaller (see Fig.~\ref{Fig:transition, different Rc}). As
nonlinear oscillating bubbles tend to form stable pairs without
coalescing, i.e. the bubbles remain at a certain equilibrium
distance from each other \cite{doi01,har01,pel04,yos11}, we can
expect that an increase in the number of bubbles will also lead to
an increase in the cluster size and therefore to a lower $P_a$ at
transition.

The dependence of the pressure amplitude for clusters merging on
the bubble size is non-monotonic. In a range from $R_0$ = 3 to
15~$\mu$m, the clusters requiring a lower $\bar{P_a}$ to escape
from the pit region are those with $R_0$ = 10~$\mu$m (see
Fig.\ref{Fig:transition, different R0}). Therefore we can assume
that these bubbles will also be the ones to initiate the merging.
Nevertheless, the inclusion of inertia and drag, excluded from the
present approach, could shift this minimum towards lower values of
$R_0$ and it should be addressed in a future paper.

\subsection{Radical production}

In order to investigate the effects of the Bjerknes forces on the
radical production, we consider a cluster with $R_c$ = 100~$\mu$m,
$N$ = 100, $d$ = 1000~$\mu$m, $R_0$ = 3~$\mu$m, driven at $f$ =
200~kHz and $P_a$ = 270~kHz. In Fig.~\ref{Fig:R(t),T(t) with and
wo Bjerknes} we show the radial and thermal evolution of a bubble
of such a cluster. Including the Bjerknes forces (solid line) has
the same effect of adding some damping to the system, as it leads
to a lower expansion of the bubble \cite{yas08} and therefore to a
lower temperature at collapse respect to the case where Bjerknes
forces are not included (solid-dashed line). As the radical
production is related to the peak temperature through Arrhenius
law, neglecting the Bjerknes forces induces a huge overestimate of
the produced radicals (see Fig.~\ref{Fig:OH(t) with and wo
Bjerknes}). Moreover, the Bjerknes forces reduce the
eigenfrequency of the bubble (see Fig.~\ref{Fig:R(t),T(t) with and
wo Bjerknes}), and therefore induces a further reduction of the
radical production, due to the lower number of collapses per unit
time respect to the isolated bubble. From the theoretical point of
view this reduction of the resonance frequency can be predicted
using the standard approach for the calculation of the linear
resonance frequency of a bubble \cite{ple77,ple77b}. We rewrite
the modified Rayleigh-Plesset equation (\ref{Eq:modif R-Plesset
eqn}) by neglecting the effects of liquid compressibility, surface
tension and viscosity, under the hypothesis of linear
oscillations. Considering just the effect of the other bubbles of
the same cloud the new linear frequency will be such to satisfy $
\omega_0^2 = 3 \kappa p_\infty / [\rho R_0^2 (1+3/2 NR_0 /R_c)] $.
With $N$ varying between 10 and 100, $NR_0/R_c$ varies between 1
and 10. For $R_0 = 10~\mu$m, we compute $f_0$ = 326, 304, 206, and
82~kHz, respectively for $NR/R_c = 0$ (single bubble), 1, 10, and
100. Given the resonance frequency of the pit $f_p$ = 143~kHz and
the driving frequency $f$ = 200~kHz, in the linear regime, i.e. at
low driving amplitude, an attractive pit-cluster force is
expected, in agreement with what we found. However, due to
nonlinearities, at high driving amplitude, the pit-cluster force
can become repulsive \cite{met97,yos11} (see Fig.~\ref{Fig:forces,
delta=R0}).
 In Fig.~\ref{Fig:R(t),T(t) with and wo Bjerknes} and in
Fig.~\ref{Fig:OH(t) with and wo Bjerknes} we also show the
dynamic, thermal, and chemical evolution of the same bubble driven
at $P_a$ = 160~kPa, without the Bjerknes forces. This driving
amplitude corresponds to the "effective pressure" that we
calculated in our previous work for the two pits case, just before
the merging occurred (see Fig. 21 in Ref.\cite{fer13}). In that
case the effective pressure was extracted from the experimental
data by neglecting the Bjerknes forces and using the bubbles as a
pressure sensor through their recorded dynamics. Although there
are some differences between the radial evolution curve of a
bubble driven at $P_a$ = 270~kPa undergoing Bjerknes forces and an
isolated bubble driven at $P_a$ = 160~kPa, the maximum and the
minimum radius correspond, thus providing consistency between the
present work and Ref.~\cite{fer13}.

The interaction with the other bubbles strongly influences the
radical production even before the inception of the motion. In
Fig.~\ref{Fig:d-OH, different Rc} we show the maximum number of OH
radicals produced per cycle in one bubble of cluster~1 as function
of the distance between the pits. When the distance between the
pits decreases, so does the radical production, because the
interaction with the neighboring bubbles becomes stronger,
therefore damping the oscillations and decreasing the temperature
at collapse. For the same reason, for a fixed number of bubbles,
smaller clouds have lower radical production (Fig.~\ref{Fig:d-OH,
different Rc}), as the bubbles are closer. Similarly, for a fixed
cluster size, raising the number of bubbles reduces the chemical
production (see Fig.~\ref{Fig:d-OH, different N}). However, in the
range between 50 and 100 bubbles, there is a local maximum at
$N\sim 70$. This could provide an explanation to the experimental
observation that, above a certain threshold, a further increase of
the applied acoustic power does not enhance the radical production
\cite{fer13}.

\section{Conclusions}

In the present work we theoretically studied the interactions
between bubbles clusters generated from ultrasonically driven
silicon etched micropits \cite{fer13}. We addressed the transition
between the three different behaviors observed in
Ref.~\cite{fer13} at increasing acoustic power: clusters sitting
upon their own pit, clusters pointing towards each other and
clusters migrating towards the center point of the pits array. We
considered each cluster as a point object and we examined the
secondary Bjerknes forces acting upon it. These forces depend on
the displacement of the cluster from the pit. While the
cluster-cluster force is always attractive, in the considered
parametric range, at high driving the pit-cluster force can also
become repulsive at some points very close the pit, thus favoring
the inception of the motion. Given the driving frequency and the
size of the bubbles and the pits, this is in contrast with the
predictions of the linear theory and it has to be ascribed to the
nonlinearity of the phenomenon \cite{ogu90,met97,yos11}. We found
that there always exists a barrier, generally coinciding with the
maximum attractive force of the pit, that needs to be overcome for
cluster-merging to take place. This barrier is located at a
distance of the order of the cluster radius, in agreement with
experimental observations. The $P_a$ required for the cluster to
escape the trapping force of the pit is consistent both with the
measured ones and with the effective pressures that we extracted
from the bubble dynamics in our previous work \cite{fer13}. As the
distance between the pits increases, the $P_a$ for merging also
increases, up to a certain limiting distance, where the cluster
cannot escape from the pits. For practical purposes, this could be
regarded as an optimal distance between the pits for efficient
sonochemical reactors design, where the number of the pits (i.e.
of the bubbles) should be the maximum possible per unit area, but
still avoiding the merging as it lowers the radical production
\cite{fer13} and enhances the erosion of the reactor walls
\cite{fer12b}.

We showed that the key to the transition to merging clusters
relies on the influence of the nonlinear bubble oscillations and
not on the increasing number of bubbles at increasing powers, as
the $P_a$ for merging decreases when $N$ increases, but only until
a certain threshold, after which it basically becomes constant.
However, an indirect influence of the number of bubbles can be
present, because the $P_a$ for transition decreases with the size
of the cluster. As strongly driven bubbles tend to form stable
pairs rather than to coalesce \cite{met97}, we can expect that
increasing the number of generated bubbles will also increase the
clusters radii. This dependence of the cluster dimension on the
number of bubbles, as well as inertia and drag experienced by
translating bubbles, should be addressed in future works, for a
more complete understanding of the phenomenon.

We also showed that the bubble size has an influence on the
driving pressure required for merging, which presents a minimum
for clusters with bubbles of 10~$\mu$m. We can therefore expect
that the merging and the consequent generated flow are first
initiated by the bubbles of this size, with some shift towards the
smaller bubbles once that also inertia and drag come into play.

Finally, we examined the sonochemical production and we found that
neglecting the Bjerknes forces will lead to a huge overestimate of
the number of radicals produced. This happens because the
interaction with the neighbors dampens the oscillations of the
bubble, reducing the temperature at collapse and also the
resonance frequency. Since these interactions exhibit an inverse
proportionality with the distance between the bubbles, smaller
size of the clusters and shorter distances between the pits as
well as higher number of bubbles strongly decrease the radical
production, even before the merging takes place. This could be the
key to explain the experimental observation that increasing the
power after a certain threshold does not improve the sonochemical
production. For practical purposes, the efficiency of a
sonochemical reactor could benefit from a medium power operating
condition instead of high power and a distance between the pits
substantially larger, in order to prevent cluster merging, which
reduces the chemical yield.

\begin{acknowledgments}
We acknowledge innumerous discussions with Andrea Prosperetti over
the years, from whom we learnt tremendously. We moreover
acknowledge Technology Foundation STW and the Nederlandse
Organisatie voor Wetenschappelijk Onderzoek (NWO) for financial
support.
\end{acknowledgments}

\newpage

\clearpage

\bibliographystyle{jasanum}

\newpage

\begin{figure}[htbp]
\centering
\includegraphics[width=0.8 \textwidth]{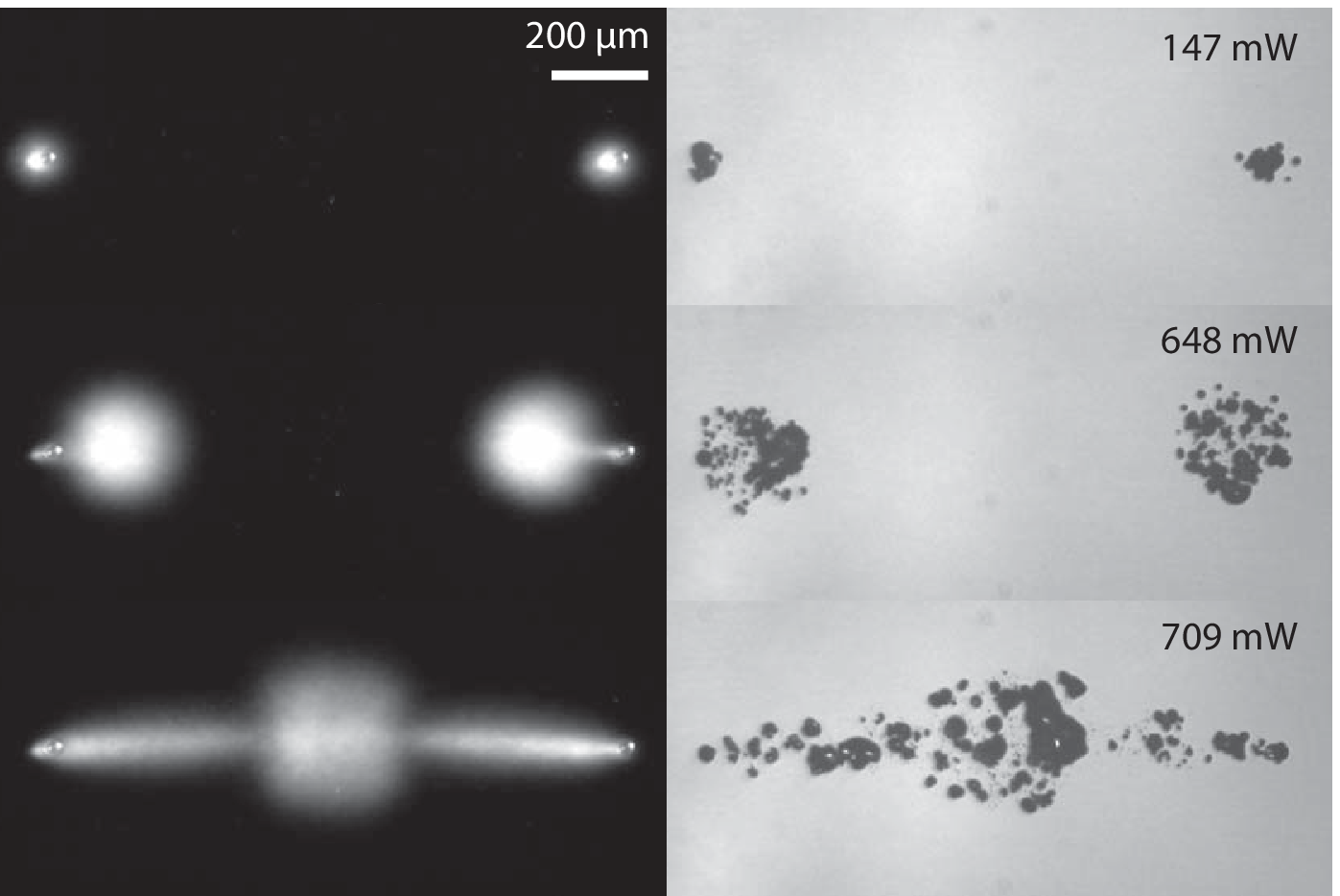}
\caption{Example of the three different behaviors of the clusters
observed in experiments, at increasing applied power: clusters
close to the pits from which their bubbles originated (behavior~1,
top), clusters pointing toward the center (behavior~2, middle),
and clusters migrating toward the center (behavior~3, bottom). The
left column was recorded at normal speed and represents therefore
a time average, while the right column shows single snapshots
taken with 7~ns exposure time. The experimental conditions are the
same described in Ref.~\cite{fer13}.}
 \label{Fig:clusters behavior, experiment}
\end{figure}

\begin{figure}[htbp]
\centering
\includegraphics[width=0.9 \textwidth]{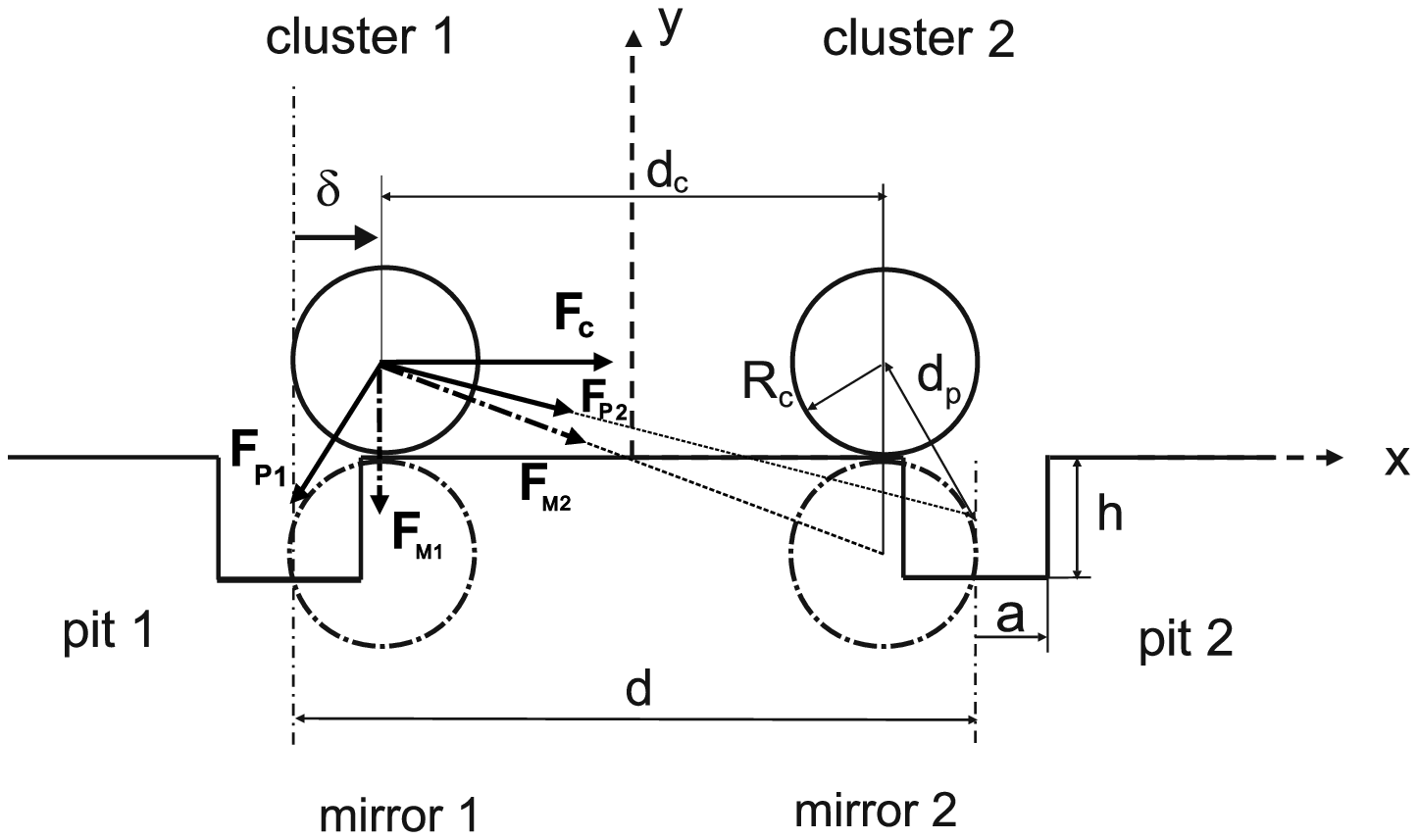}
\caption{Sketch of the secondary Bjerknes forces acting upon the
bubble and introduction of our employed notation. $F_c$ is the
force from cluster~2, $F_{p1,2}$ are forces from the pits and
$F_{m1,2}$ are the forces from the mirror clusters.}
\label{Fig:Sketch forces}
\end{figure}

\begin{figure}[htbp]
\centering
\includegraphics[width=0.9 \textwidth]{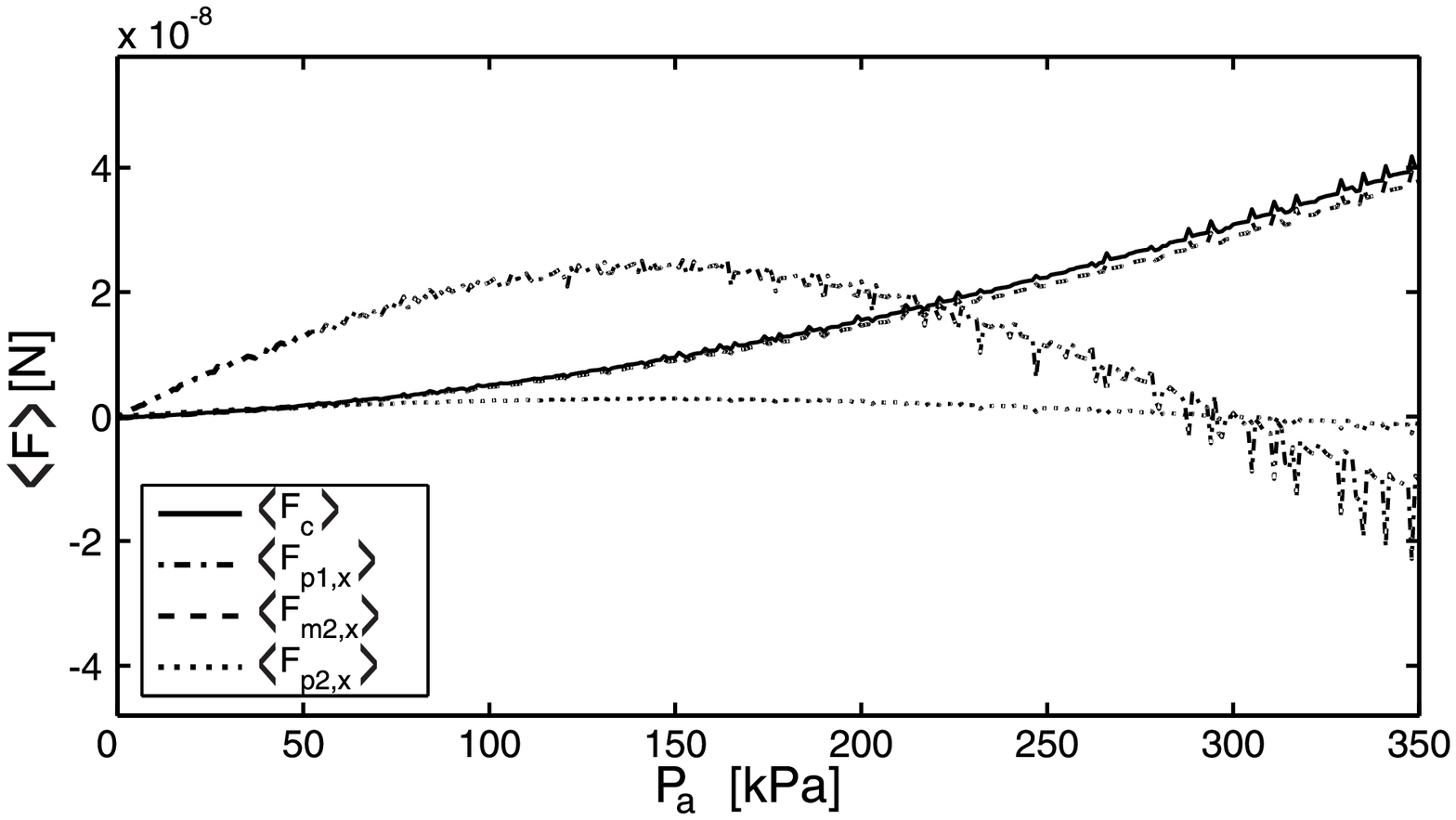} \caption{Forces
acting upon a bubble of cluster~1 with an initial displacement
$\delta = R_0$ respect to its unperturbed condition. Parameters:
$R_c$ = 100~$\mu$m, $N$ = 100, $d$ = 1000~$\mu$m, $R_0 $ =
10~$\mu$m, $f$ = 200~kHz. The cluster-cluster forces are always
attractive, while the pit-cluster forces become repulsive at high
driving, thus favoring the inception of the motion.}
 \label{Fig:forces, delta=R0}
\end{figure}

\begin{figure}[htbp]
\centering
\includegraphics[width=0.9\textwidth]{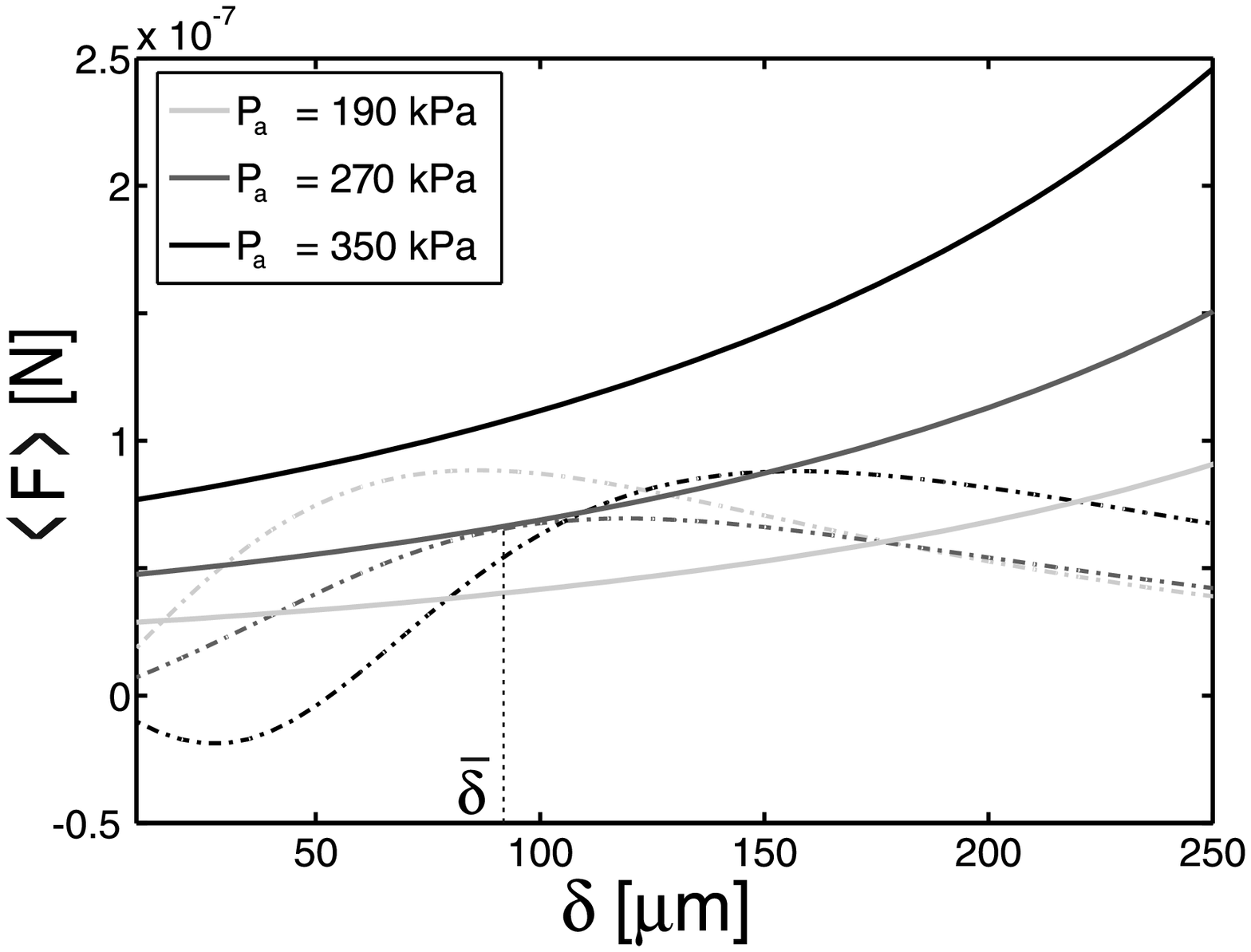}
\caption{Average forces $\langle F_x^* \rangle$ (solid) and
$\langle F_{p1,x}\rangle$ (dash-dot) acting upon a bubble of
cluster~1, over one acoustic cycle, as function of $\delta$ for
$R_0$ = 10~$\mu$m, $N$ = 100, $R_c$ = 100~$\mu$m, $d$ =
1000~$\mu$m, $f$ = 200~kHz. In lit-gray, low pressure case:
$\langle F_x^* \rangle$ is too weak to pass the barrier
constituted by the restoring force $\langle F_{p1,x}\rangle$; the
cluster remains attached to its pit, with a small displacement
given by the first intersection of the two curves. In black, high
pressure case: $\langle F_x^* \rangle > \langle F_{p1,x}\rangle$,
and the two clusters merge. In mid-gray, critical case, defining
the transition between both behaviors. $\bar{\delta}$ represents
the maximum displacement of the cluster before the transition. }
 \label{Fig:transition}
\end{figure}

\begin{figure}[htbp]
\centering
\includegraphics[width=0.9\textwidth]{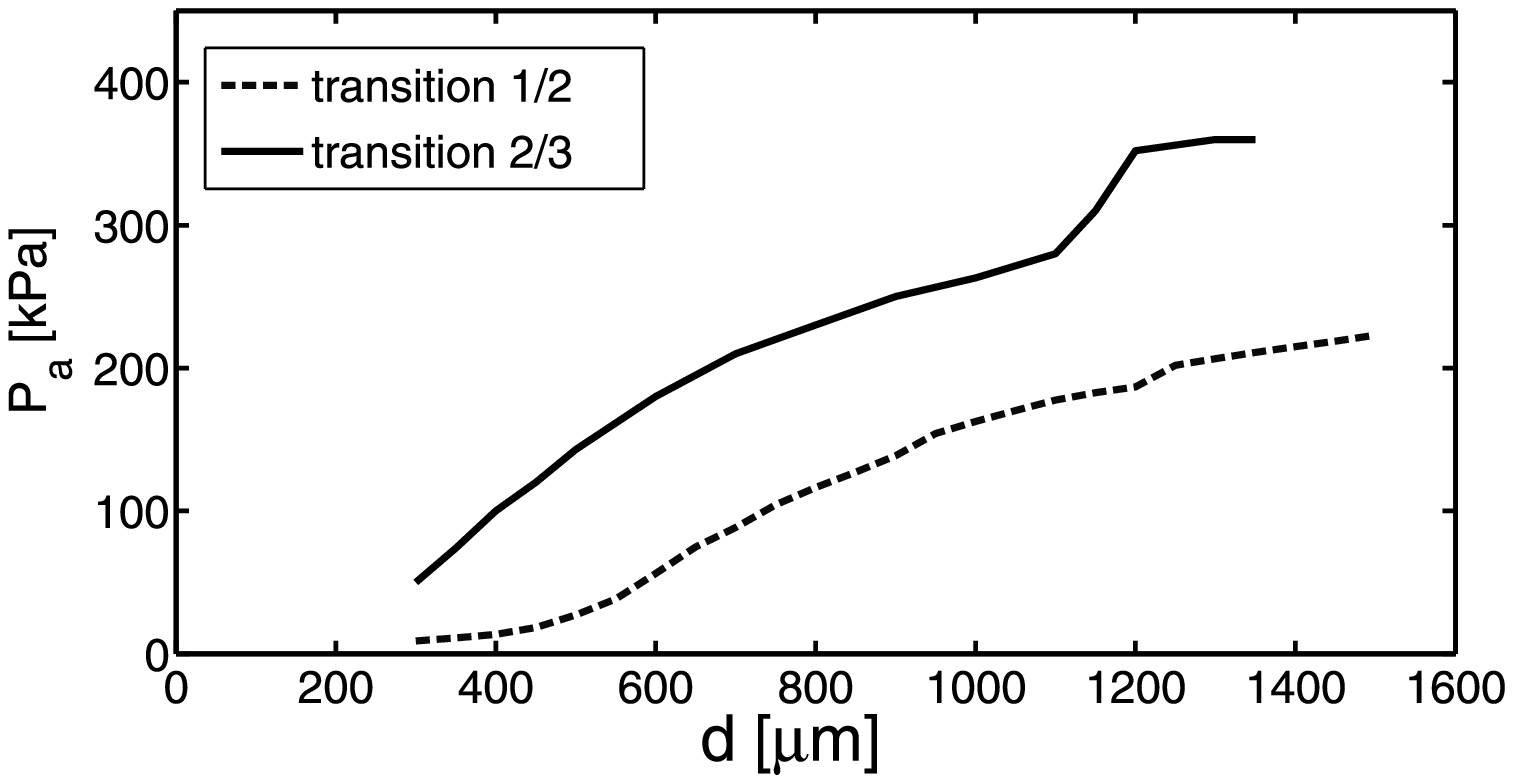}
\caption{Calculated driving pressure for transition from
behavior~1 to 2, i.e. the inception of the motion (dash) and from
behavior~2 to 3, i.e. the merging of the clusters (solid), as
function of the distance between the pits for a cluster with $R_0$
= 10~$\mu$m, $N$ = 100, $R_c$ = 100~$\mu$m, driven at $f$ =
200~kHz. When the pits are too far apart ($d \gtrsim 1350 \
\mu$m), no merging is possible. }
 \label{Fig:d-Pa, transition}
\end{figure}

\begin{figure}[htbp]
\centering
\includegraphics[width=0.9\textwidth]{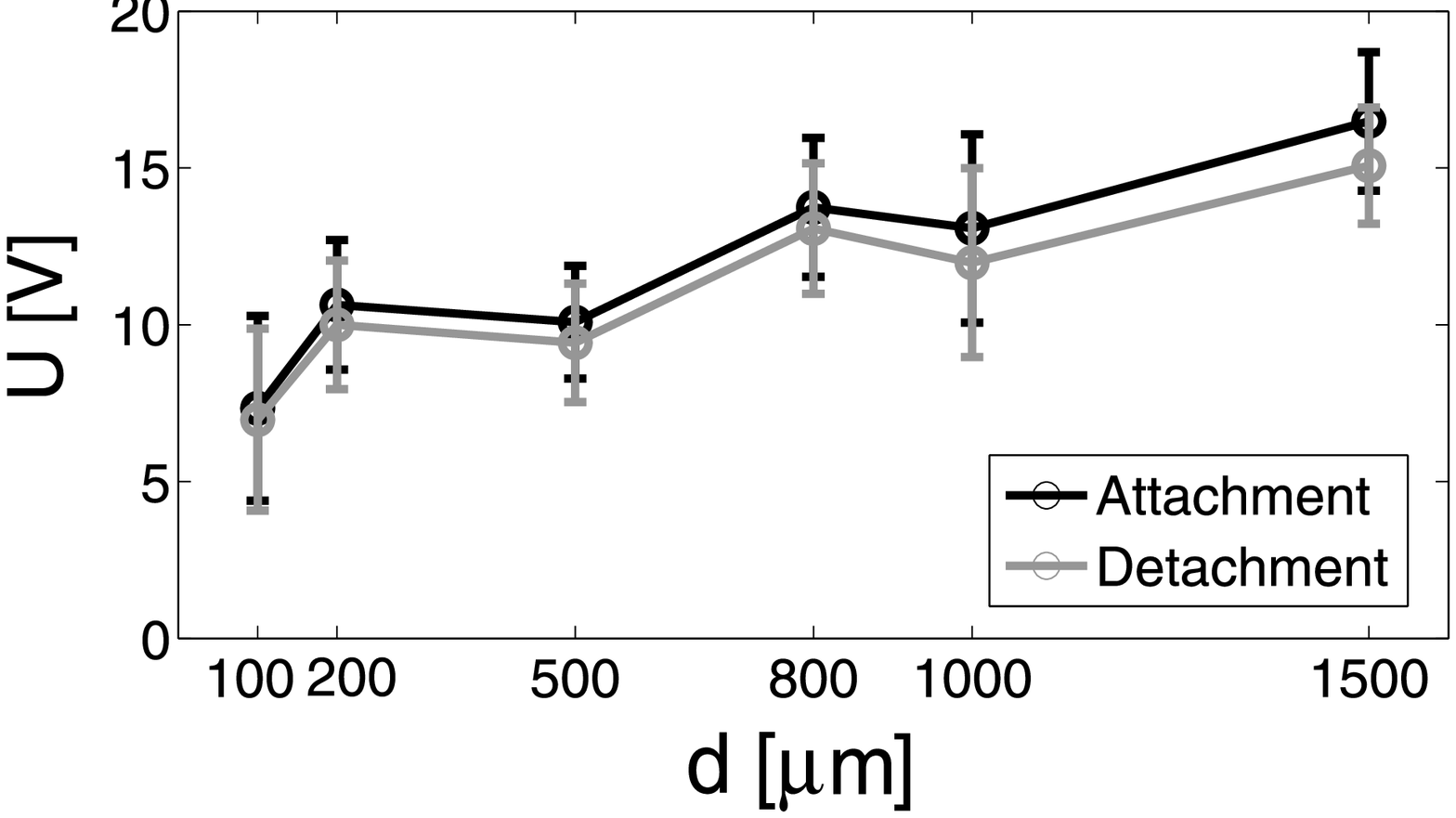}
\caption{Electric potential applied to the piezo when the clusters
start to merge, at increasing power (black), and when they detach
going back to their own pits, at decreasing power (gray), as
function of the distance between the pits, in the experiment
described in Ref.\cite{fer13}. Increasing voltage corresponds to
increasing $P_a$. For $d > 1500 \mu$m no merging was observed, no
matter $P_a$. }
 \label{Fig:transition experiment}
\end{figure}

\begin{figure}[htbp]
\centering
\includegraphics[width=0.9\textwidth]{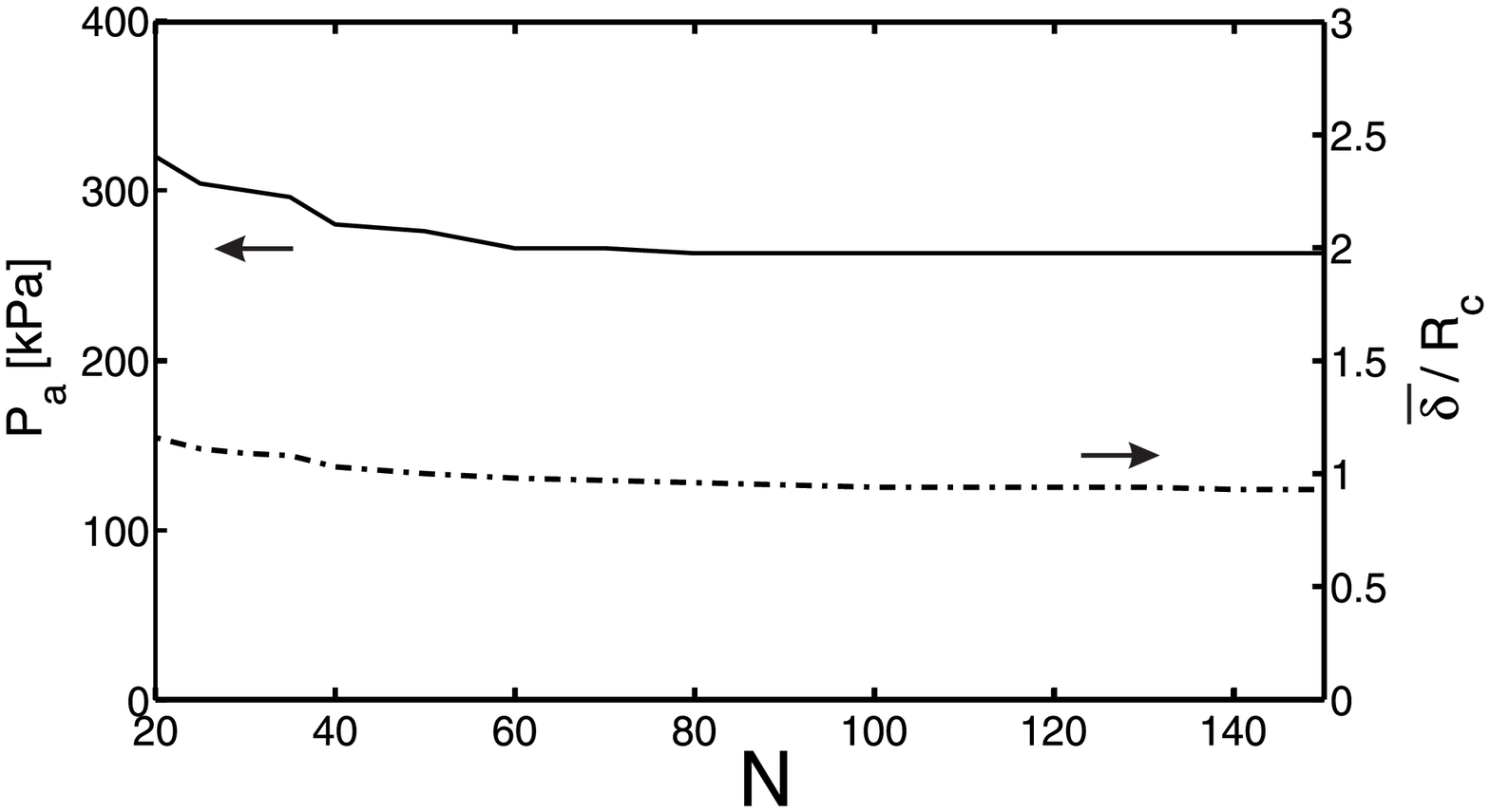}
\caption{Driving pressure (solid) and maximum displacement of the
cluster (dash-dot) at transition from behavior~2 to behavior~3, as
function of the number of bubbles. Parameters: $R_0$ = 10~$\mu$m,
$R_c$ = 100~$\mu$m, $d$ = 1000~$\mu$m, $f$ = 200~kHz. For $N > 50$
they become both almost invariant to a further increase of $N$. }
 \label{Fig:transition, different N}
\end{figure}

\begin{figure}[htbp]
\centering
\includegraphics[width=0.9\textwidth]{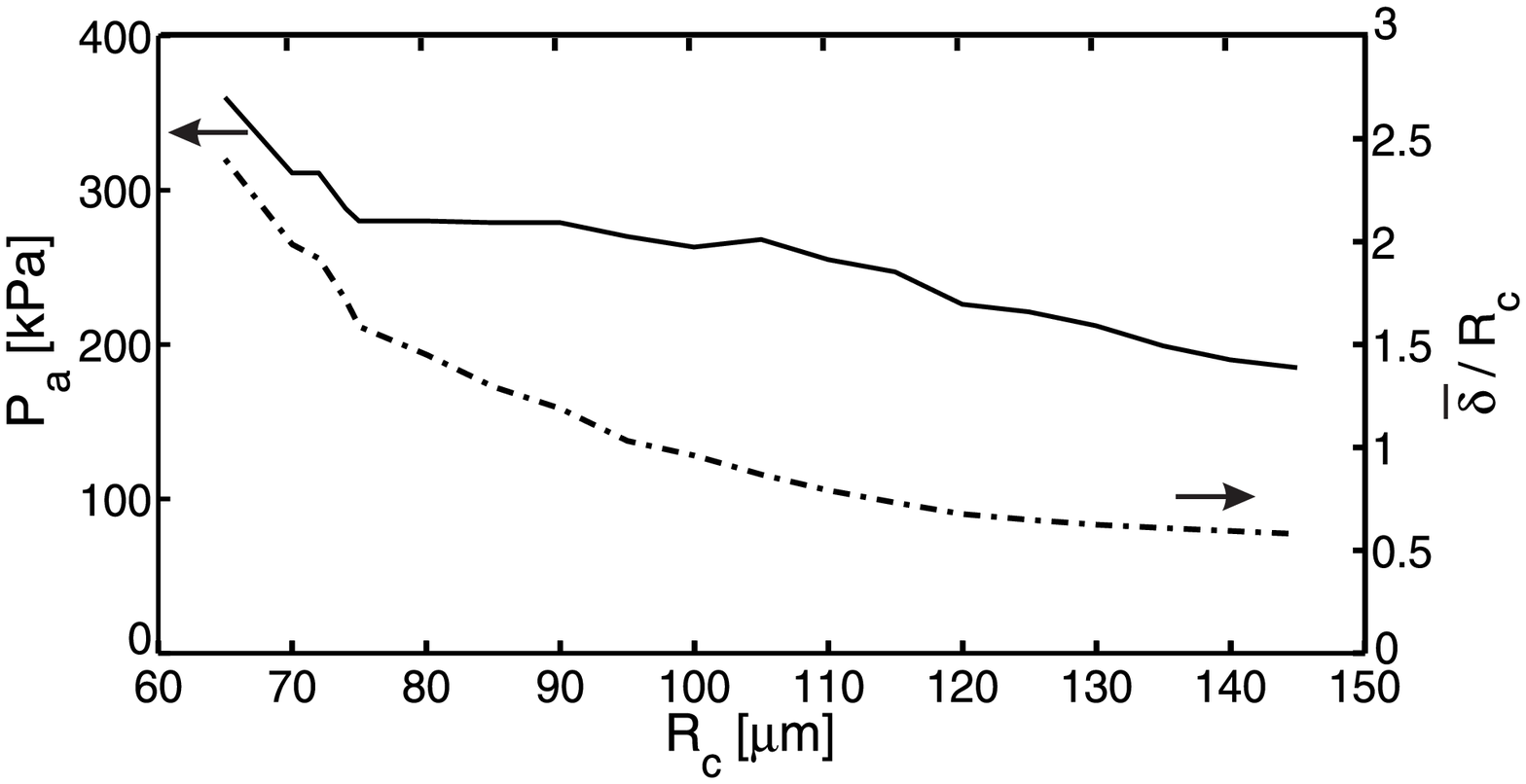}
\caption{Driving pressure (solid) and maximum displacement of the
cluster (dash-dot) at transition from behavior~2 to behavior~3, as
function of the radius of the cluster. Parameters: $R_0$ =
10~$\mu$m, $N$ = 100, $d$ = 1000~$\mu$m, $f$ = 200~kHz. For bigger
clusters the $P_a$ required for merging is lower, as well as the
maximum displacement of the cluster.}
 \label{Fig:transition, different Rc}
\end{figure}

\begin{figure}[htbp]
\centering
\includegraphics[width=0.9\textwidth]{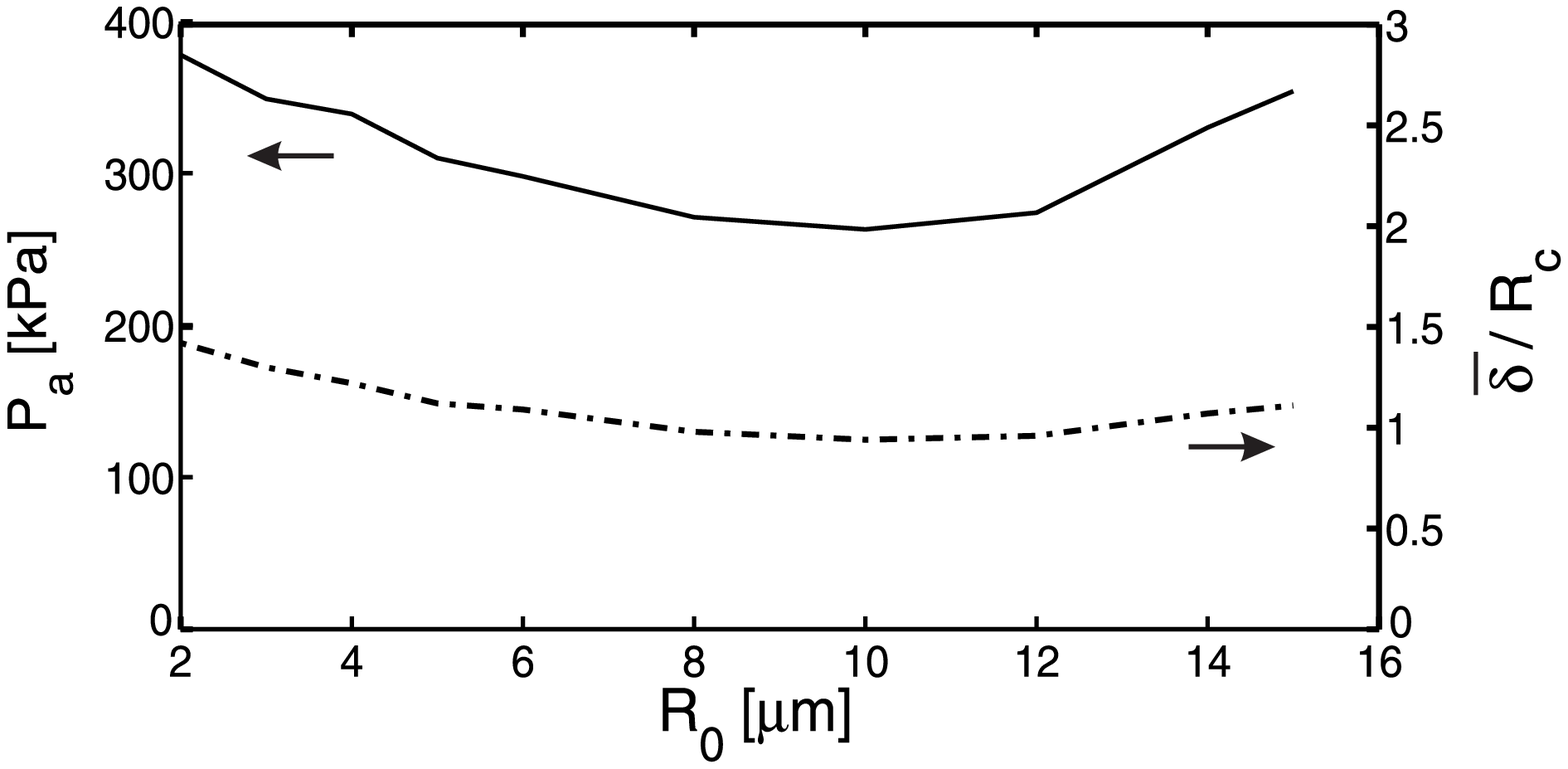}
\caption{Driving pressure (solid) and maximum displacement of the
cluster (dash-dot) at transition from behavior~2 to behavior~3, as
function of the radius of the bubbles. Parameters: $R_c$ =
100~$\mu$m, $N$ = 100, $d$ = 1000~$\mu$m, $f$ = 200~kHz. The $P_a$
required for merging presents a minimum with bubbles of 10~$\mu$m,
as well as the maximum displacement, which is however almost
invariant with $R_0$. }
 \label{Fig:transition, different R0}
\end{figure}

\begin{figure}[htbp]
\centering
\includegraphics[width=0.9\textwidth]{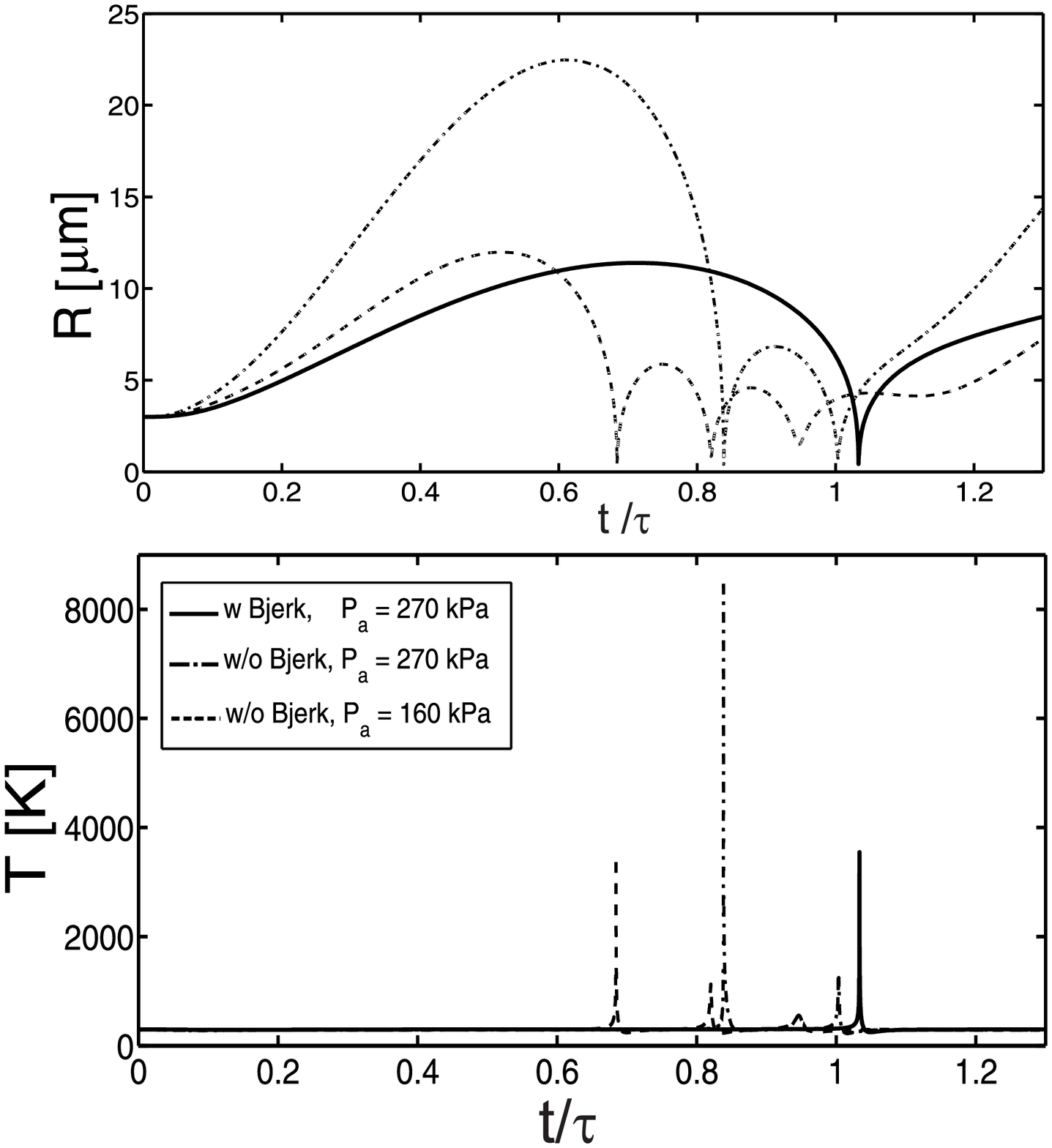}
\caption{Radius vs time (top) and temperature vs time (bottom)
curves, for a bubble belonging to cluster~1. The driving amplitude
corresponds to the calculated one for the transition between
behavior~2 and behavior~3, both including (solid) and disregarding
(dash-dot) the Bjerknes forces for $P_a$ = 270~kPa. The dashed
lines correspond to an isolated bubble, driven at $P_a$ = 160~kPa,
that equals the "effective pressure" deduced from the bubble
dynamics disregarding Bjerknes forces, just before transition in
Ref.~\cite{fer13}. Parameters: $R_0$ = 3~$\mu$m, $R_c$ =
100~$\mu$m, $N$ = 100, $f$ = 200~kHz, $d$ = 1000~$\mu$m, $a$ =
15~$\mu$m, $h$ = 10~$\mu$m. } \label{Fig:R(t),T(t) with and wo
Bjerknes}
\end{figure}

\begin{figure}[htbp]
\centering
\includegraphics[width=0.9\textwidth]{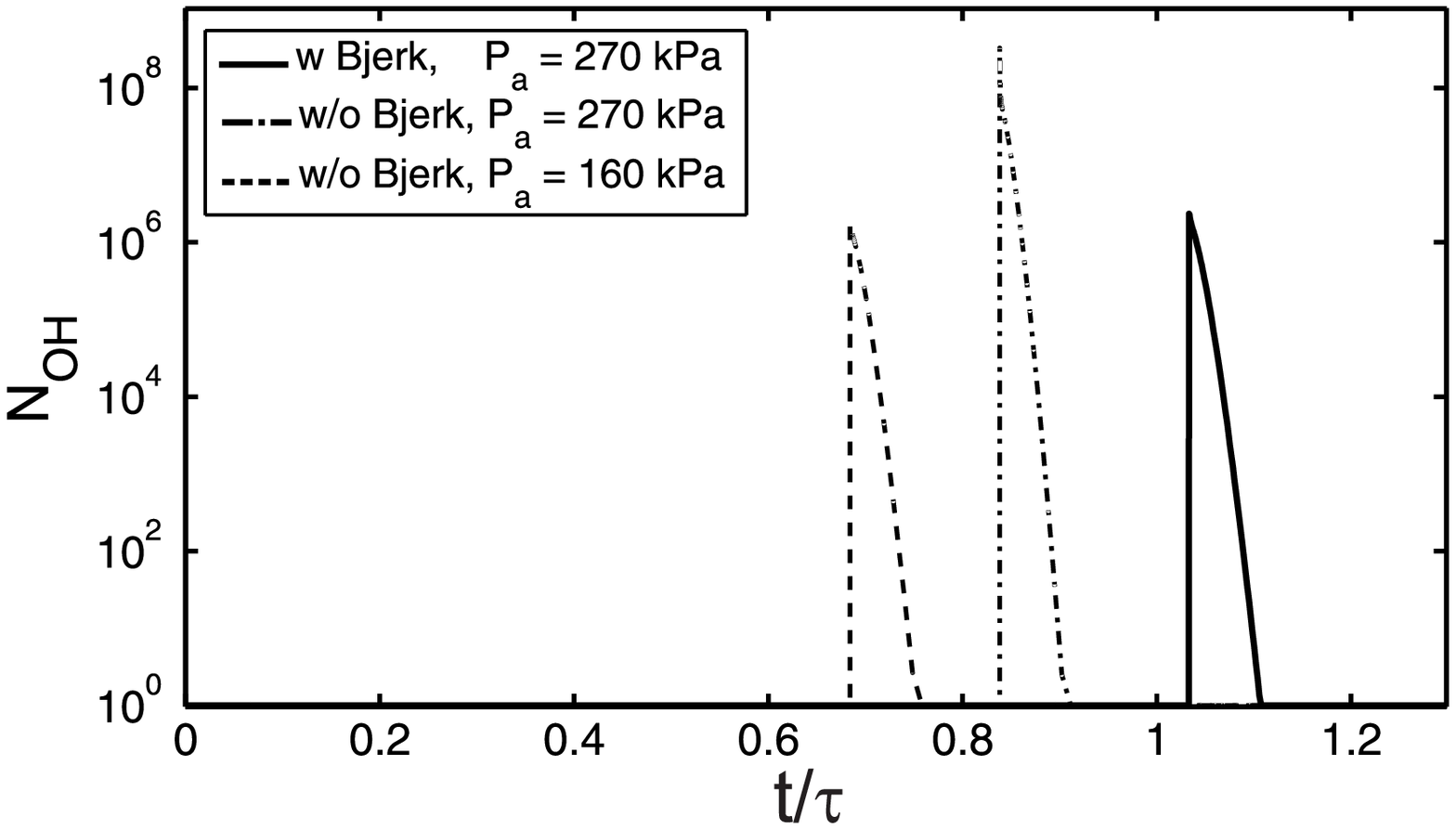}
\caption{Number of OH radicals produced as function of time, for a
bubble belonging to cluster~1 driven at $P_a$ = 270~kPa, with
(solid) and without (dash-dot) Bjerknes forces, and at $P_a$ =
160~kPa without Bjerknes forces (dash). Driving conditions and
dimensions as in Fig.~\ref{Fig:R(t),T(t) with and wo Bjerknes}.}
\label{Fig:OH(t) with and wo Bjerknes}
\end{figure}

\begin{figure}[htbp]
\centering
\includegraphics[width=0.9\textwidth]{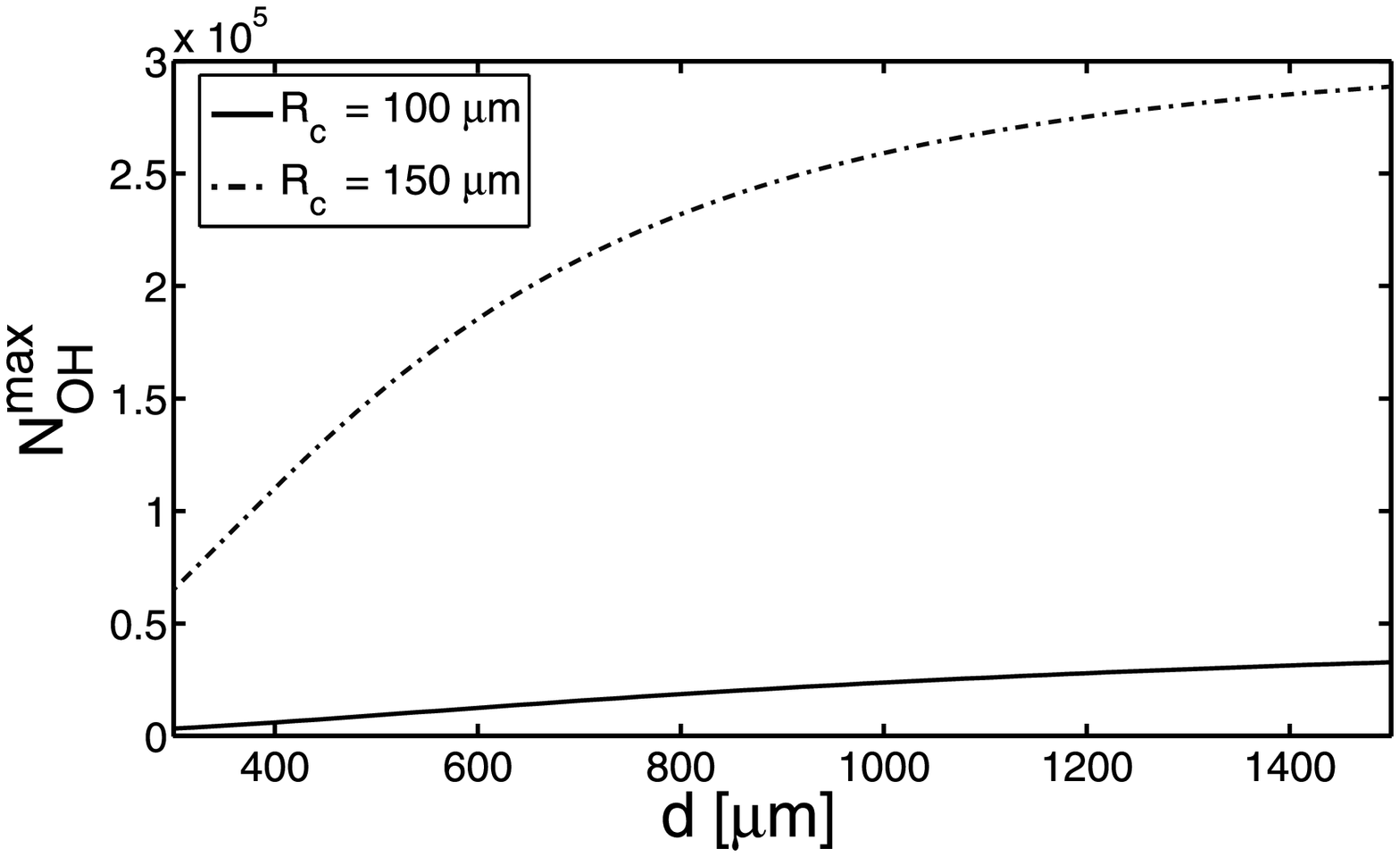}
\caption{ Maximum number of OH radicals in one acoustic cycle for
a bubble of cluster~1 in the unperturbed position ($\delta = 0 \
\mu m$) as function of the distance between the pits, for
different cluster sizes (as indicated in the legend). Parameters:
$N$ = 100, $R_0 $ = 3~$\mu$m, $f$ = 200~kHz, $P_a$ = 170~kPa. }
 \label{Fig:d-OH, different Rc}
\end{figure}

\begin{figure}[htbp]
\centering
\includegraphics[width=0.9\textwidth]{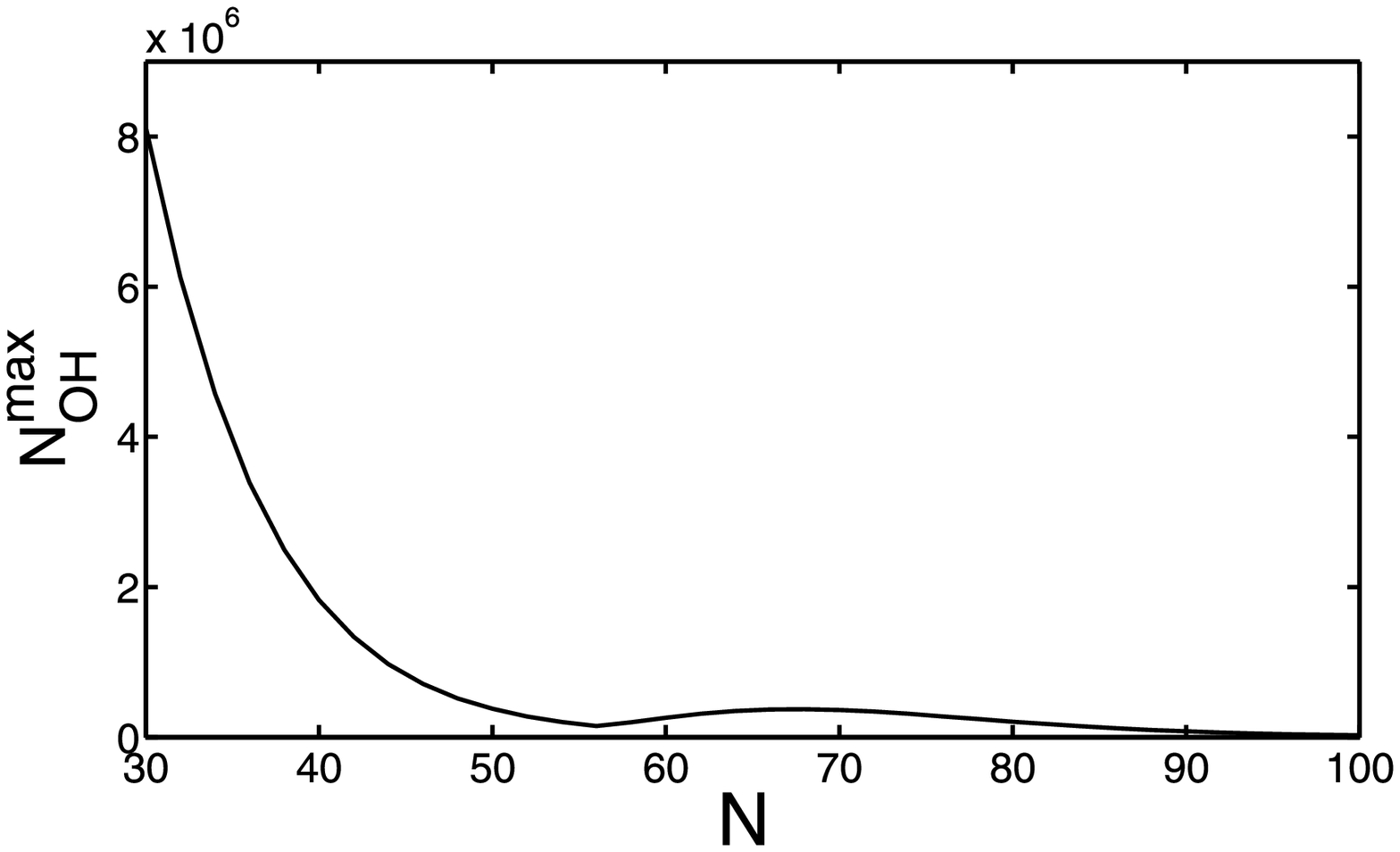}
\caption{ Maximum number of OH radicals in one acoustic cycle for
a bubble of cluster~1 in the unperturbed position ($\delta = 0 \
\mu m$) as function of the number of bubbles. Parameters: $R_c$ =
100~$\mu$m, $d$ = 1000~$\mu$m, $R_0 $ = 3~$\mu$m, $f$ = 200~kHz,
$P_a$ = 170~kPa.}
 \label{Fig:d-OH, different N}
\end{figure}

\end{document}